\documentclass{article}%
\usepackage{amsfonts}
\usepackage{amsmath}
\usepackage{amssymb}
\usepackage{graphicx}%
\providecommand{\U}[1]{\protect\rule{.1in}{.1in}}
\providecommand{\U}[1]{\protect\rule{.1in}{.1in}}
\newtheorem{theorem}{Theorem}

\newtheorem{corollary}[theorem]{Corollary}

\newtheorem{example}[theorem]{Example}

\newtheorem{remark}[theorem]{Remark}

\newenvironment{proof}[1][Proof]{\noindent\textbf{#1.} }{\ \rule{0.5em}{0.5em}}
\numberwithin{equation}{section}
\begin{document}

\title{St\"{a}ckel transform of Lax equations}
\author{Maciej B\l aszak\\Faculty of Physics, Division of Mathematical Physics, A. Mickiewicz University\\Umultowska 85, 61-614 Pozna\'{n}, Poland\\blaszakm@amu.edu.pl
\and Krzysztof Marciniak\\Department of Science and Technology \\Campus Norrk\"{o}ping, Link\"{o}ping University\\601-74 Norrk\"{o}ping, Sweden\\krzma@itn.liu.se}
\maketitle

\begin{abstract}
We construct Lax pairs for a wide class of St\"{a}ckel systems by applying the
multi-parameter St\"{a}ckel transform to Lax pairs of a suitably chosen
systems from the seed class. For a given St\"{a}ckel system, the obtained set
of non-equivalent Lax pairs is parametrized by an arbitrary function.

\end{abstract}


\section{Introduction}

In this paper we show how to construct Lax pairs for St\"{a}ckel systems
generated by spectral curves of the form%
\begin{equation}
\sigma(\lambda)+%
{\displaystyle\sum\limits_{j=1}^{n}}
\tilde{H}_{j}\lambda^{\gamma_{j}}=\frac{1}{2}f(\lambda)\mu^{2}\text{, \ }
\label{Si}%
\end{equation}
where $\gamma_{i}\in\mathbb{N}$ with the normalization $\gamma_{1}>\gamma
_{2}>\cdots>\gamma_{n}=0$ and where $\sigma,$ $f$ are Laurent polynomials in
$\lambda$. Taking \thinspace$n$ copies of this curve at points $(\lambda
_{i},\mu_{i})$ we obtain a system of $n$ linear equations for $\tilde{H}_{j}$
and its solution yield $n$ commuting, in the sense of canonical Poisson
bracket $\left\{  \lambda_{i},\mu_{j}\right\}  =\delta_{ij}$, St\"{a}ckel
Hamiltonians $\tilde{H}_{j}(\lambda,\mu)$. Such St\"{a}ckel systems are fairly
general. Each choice of the constants $\gamma=\left\{  \gamma_{1}%
,\cdots,\gamma_{n}\right\}  $ fixes the St\"{a}ckel matrix $S_{ij}=\lambda
_{i}^{\gamma_{j}}$ in (\ref{Si}). Unless we specify the functions $\sigma$ and
$f$ the curve (\ref{Si}) yields a family of systems that we will call a
$\gamma$-class. Choosing the constants $\gamma_{j}$ as $\gamma_{j}=n-j$,
$j=1,\ldots,n$ yields the special $\gamma$-class
\begin{equation}
\sigma(\lambda)+\sum_{j=1}^{n}H_{j}\lambda^{n-j}=\frac{1}{2}f(\lambda)\mu
^{2},\ \ \label{Bi}%
\end{equation}
which is called the Benenti class \cite{Ben1, Ben2, blasz2005} or the seed class.

In paper \cite{blasz2012} we demonstrated how to generate the Hamiltonians
$\tilde{H}_{i}$ of the $\gamma$-class (\ref{Si}) as a multi-parameter
St\"{a}ckel transform \cite{hiet, boy, ts, ts2, serg2008, blasz2011,
blasz2017} of Hamiltonians $H_{i}$ from the class (\ref{Bi}). Also, in the
recent paper \cite{blasz2018} the authors found Lax pairs $(L(\lambda
),U_{i}(\lambda))$, $i=1,\ldots,n$ for equations generated by the Hamiltonians
given by (\ref{Bi}). Combining the ideas of these papers we will find Lax
pairs $(\tilde{L}(\lambda),\tilde{U}_{i}(\lambda))$, $i=1,\ldots,n$ for all
the Hamiltonian systems of a given $\gamma$-class (\ref{Si}). In order to do
this we will make an appropriate extension of Hamiltonians from both classes
(\ref{Si}) and (\ref{Bi}) by a number of parameters $\alpha_{i}$ and
$\tilde{\alpha}_{i}$ so that the extended systems will be related by a
multi-parameter St\"{a}ckel transform. Applying this St\"{a}ckel transform to
the Lax pairs $(L(\lambda,a),U_{i}(\lambda,a))$ of the parameter-dependent
systems from the seed class, we will obtain the Lax pairs $(\tilde{L}%
(\lambda,\tilde{a}),\tilde{U}_{i}(\lambda,\tilde{a}))$ of the
parameter-dependent Hamiltonians of $\gamma$-class. Finally, by setting all
the parameters to zero we will obtain the sought Lax pairs $(\tilde{L}%
(\lambda),\tilde{U}_{i}(\lambda))$ for (\ref{Si}).

The paper is organized as follows. In Section \ref{S1} we remind the main
concepts of the multi-parameter St\"{a}ckel transform. In Section \ref{S2} we
prove Theorem \ref{T1} describing how Lax pairs transform under the
St\"{a}ckel transform. Section \ref{S3} is devoted to Hamiltonian systems and
their Lax representations $(L(\lambda),U_{i}(\lambda))$ from the seed class
(\ref{Bi}). In Section \ref{S4} we remind the St\"{a}ckel transform relating
systems from the seed class (\ref{Bi}) and these from arbitrary $\gamma$-class
(\ref{Si}). Finally, in Section \ref{S5}, we apply the results of Section
\ref{S2} and Section \ref{S4}\ in order to construct the Lax pairs for the
St\"{a}ckel systems from $\gamma$-classes (\ref{Si}). The paper is furnished
with few examples.

\section{St\"{a}ckel transform of integrable Hamiltonian systems\label{S1}}

Let us consider a Liouville integrable Hamiltonian system on a $2n$%
-dimensional Poisson manifold $(M,\pi)$ defined by $n$ Hamiltonians
$h_{i}:M\rightarrow\mathbb{R}$ on $M,$ each depending on $k\leq n$ parameters
$\alpha_{1},\dots,\alpha_{k}$ so that
\begin{equation}
h_{i}=h_{i}(\xi,\alpha_{1},\dots,\alpha_{k}),\quad i=1,\dots,n \label{h}%
\end{equation}
where $\xi\in M$. From $n$ functions in (\ref{h}) we choose $k$ functions
$h_{s_{i}}$, $i=1,\ldots,k$, where $\{s_{1},\dots,s_{k}\}$ $\subset
\{1,\dots,n\}$. Solving (we assume it is globally possible) the system of
equations%
\begin{equation}
h_{s_{i}}(\xi,\alpha_{1},\dots,\alpha_{k})=\tilde{\alpha}_{i},\quad
i=1,\dots,k \label{hsi}%
\end{equation}
(where $\tilde{\alpha}_{i}$ is another set of $k$ parameters) with respect to
$\alpha_{i}$ yields%
\begin{equation}
\alpha_{i}=\tilde{h}_{s_{i}}(\xi,\tilde{\alpha}_{1},\dots,\tilde{\alpha}%
_{k}),\quad i=1,\dots,k, \label{alphy}%
\end{equation}
where the right-hand sides of these solutions define $k$ new functions
$\tilde{h}_{s_{i}}$ on $M$, each depending on $k$ parameters $\tilde{\alpha
}_{i}$. Let us also define $n-k$ functions $\tilde{h}_{i}$ by substituting
$\tilde{h}_{s_{i}}$ instead of $\alpha_{i}$ in $h_{i}$ for all $i\notin
\{s_{1},\dots,s_{k}\}$%
\begin{equation}
\tilde{h}_{i}=h_{i}|_{\alpha_{1}\rightarrow\tilde{h}_{s_{1}},\ldots,\alpha
_{k}\rightarrow\tilde{h}_{s_{k}}},\quad i=1,\ldots,n,\text{ \ }i\notin
\{s_{1},\dots,s_{k}\}. \label{reszta}%
\end{equation}
The functions $\tilde{h}_{i}=\tilde{h}_{i}(\xi,\tilde{\alpha}_{1},\dots
,\tilde{\alpha}_{k}),$ $i=1,\dots,n$, defined through (\ref{alphy}) and
(\ref{reszta}) are called the $k$-parameter St\"{a}ckel transform of the
functions (\ref{h}). If we perform again the St\"{a}ckel transform on the
functions $\tilde{h}_{i}$ with respect to $\tilde{h}_{s_{i}}$ we will receive
back the functions $h_{i}$ in (\ref{h}). One can prove \cite{serg2008,
blasz2012} that St\"{a}ckel transform preserves functional independence as
well as involutivity with respect to $\pi$.

The Hamiltonians $h_{i}$ yield $n$ commuting Hamiltonian systems on $M$%
\begin{equation}
\frac{d\xi}{dt_{i}}=\pi dh_{i}\equiv X_{i}\text{, }i=1,\ldots,n \label{ham}%
\end{equation}
depending on $k$ parameters $\alpha_{i}$, while $\tilde{h}_{i}$ define $n$
commuting systems
\begin{equation}
\frac{d\xi}{d\tilde{t}_{i}}=\pi d\tilde{h}_{i}\equiv\tilde{X}_{i}\text{,
}i=1,\ldots,n \label{hamtil}%
\end{equation}
depending on $k$ parameters $\tilde{\alpha}_{i}$.

Observe that as soon as we fix the values of both all $\alpha_{i}$ and all
$\tilde{\alpha}_{i}$ the relation (\ref{hsi}) defines the $(2n-k)$-dimensional
submanifold $M_{\alpha,\tilde{\alpha}}$ given by (\ref{hsi}):%
\begin{equation}
M_{\alpha,\tilde{\alpha}}=\left\{  \xi\in M:h_{s_{i}}(\xi,\alpha_{1}%
,\dots,\alpha_{k})=\tilde{\alpha}_{i}\text{, \ \ }i=1,\ldots k\right\}
\label{Maa}%
\end{equation}
or equivalently by (\ref{alphy})%
\begin{equation}
M_{\alpha,\tilde{\alpha}}=\left\{  \xi\in M:\tilde{h}_{s_{i}}(\xi
,\tilde{\alpha}_{1},\dots,\tilde{\alpha}_{k})=\alpha_{i}\text{, \ \ }%
i=1,\ldots k\right\}  \label{Mab}%
\end{equation}

\begin{remark}
\label{r}Through each point $\xi$ in $M\,\ $there passes infinitely many
submanifolds $M_{\alpha,\tilde{\alpha}}$. If we fix the values of all the
parameters $\alpha_{i}$ we can for any $\xi$ always find some values of the
parameters $\tilde{\alpha}_{i}$ so that $\xi\in M_{\alpha,\tilde{\alpha}}$ and
vice versa, if we fix $\tilde{\alpha}_{i}$, for any given $\xi$ we can find
$\alpha_{i}$ so that $\xi\in M_{\alpha,\tilde{\alpha}}$.
\end{remark}

As it follows from (\ref{hsi}), (\ref{alphy}) and (\ref{reszta}) the following
identities are valid on the whole $M$ and for all values of parameters
$\tilde{\alpha}_{i}$:
\begin{equation}
h_{s_{i}}(\xi,\tilde{h}_{s_{1}}(\xi,\tilde{\alpha}_{1},\ldots,\tilde{\alpha
}_{n}),\ldots,\tilde{h}_{s_{k}}(\xi,\tilde{\alpha}_{1},\ldots,\tilde{\alpha
}_{n}))\equiv\tilde{\alpha}_{i}\text{, \ }i=1,\ldots,k \label{id1}%
\end{equation}%
\begin{equation}
\tilde{h}_{i}(\xi,\tilde{\alpha}_{1},\ldots,\tilde{\alpha}_{n})\equiv
h_{i}(\xi,\tilde{h}_{s_{1}}(\xi,\tilde{\alpha}_{1},\ldots,\tilde{\alpha}%
_{n}),\ldots,\tilde{h}_{s_{k}}(\xi,\tilde{\alpha}_{1},\ldots,\tilde{\alpha
}_{n})),\text{ \ \ }i=1,\ldots,n\text{, \ }i\notin\{s_{1},\dots,s_{k}\}
\label{id2}%
\end{equation}
Differentiating (\ref{id1}) with respect to $\xi$ we find that \emph{on each}
$M_{\alpha,\tilde{\alpha}}$
\begin{equation}
dh_{s_{i}}=-\sum_{j=1}^{k}\frac{\partial h_{s_{i}}}{\partial\alpha_{j}}%
d\tilde{h}_{s_{j}}\text{, \ \ }i=1,\ldots,k \label{trdh1}%
\end{equation}
while differentiation of (\ref{id2}) gives that on $M_{\alpha,\tilde{\alpha}}$
we have
\begin{equation}
dh_{_{i}}=d\tilde{h}_{i}-\sum_{j=1}^{k}\frac{\partial h_{i}}{\partial
\alpha_{j}}d\tilde{h}_{s_{j}}\text{, \ \ }i=1,\ldots,n,\text{ \ }%
i\notin\{s_{1},\dots,s_{k}\}\text{\ .} \label{trdh2}%
\end{equation}
The transformations (\ref{trdh1}), (\ref{trdh2}) on $M_{\alpha,\tilde{\alpha}%
}$ can be written in a matrix form as
\begin{equation}
dh=Ad\tilde{h} \label{trdh}%
\end{equation}
where we denote $dh=(dh_{1},\ldots,dh_{n})^{T}$ and $d\tilde{h}=(d\tilde
{h}_{1},\ldots,d\tilde{h}_{n})^{T}$ and where the $n\times n$ matrix $A$ is
given by%
\begin{equation}
A_{ij}=\delta_{ij}\text{ for }j\notin\{s_{1},\dots,s_{k}\}\text{, }A_{is_{j}%
}=-\frac{\partial h_{i}}{\partial\alpha_{j}}\text{ \ for }j=1,\ldots,k
\label{A}%
\end{equation}
From the structure of the matrix $A$ it follows that
\[
\det A=\pm\det\left(  \frac{\partial h_{s_{i}}}{\partial\alpha_{j}}\right)
\]
so that $\det A\neq0$ due to our assumptions. Thus, the relation (\ref{trdh})
can be inverted yielding $d\tilde{h}=A^{-1}dh$. The relation (\ref{trdh})\ and
its inverse can be used to show the functional independence of $\tilde{h}_{i}$
for all values of $\tilde{\alpha}_{i}$ from the functional independence of
$h_{i}$ for all values of $\alpha_{i}\,$. Moreover, the same relations are
used to prove the involutivity of $\tilde{h}_{i}$ from involutivity of $h_{i}%
$. See \cite{serg2008, blasz2012} for details.

Since $X_{i}=\pi dh_{i}$ and $\tilde{X}_{i}=\pi d\tilde{h}_{i}$ we obtain from
(\ref{trdh1})-(\ref{trdh2}) that the Hamiltonian vector fields $X_{i}=\pi
dh_{i}$ and $\tilde{X}_{i}=\pi d\tilde{h}_{i}$ are on the appropriate
$M_{\alpha,\tilde{\alpha}}$ related by the following transformation%
\begin{align}
X_{s_{i}}  &  =-\sum_{j=1}^{k}\frac{\partial h_{s_{i}}}{\partial\alpha_{j}%
}\tilde{X}_{s_{j}}\text{, \ \ }i=1,\ldots,k\label{trXsi}\\
X_{_{i}}  &  =\tilde{X}_{i}-\sum_{j=1}^{k}\frac{\partial h_{i}}{\partial
\alpha_{j}}\tilde{X}_{s_{j}}\text{, \ \ }i=1,\ldots,n,\text{ \ }i\notin
\{s_{1},\dots,s_{k}\}\text{\ } \label{trXi}%
\end{align}

This means that the Hamiltonian vector fields $X_{i}$ and $\tilde{X}_{i}$ span
on each $M_{\alpha,\tilde{\alpha}}$ the same $n$-dimensional distribution and
also that the vector fields $X_{s_{i}}$ and $\tilde{X}_{s_{i}}$ span on each
$M_{\alpha,\tilde{\alpha}}$ the same $k$-dimensional subdistribution of the
above distribution. The transformation (\ref{trXsi})-(\ref{trXi}) on
$M_{\alpha,\tilde{\alpha}}$ can be written in matrix form as
\begin{equation}
X=A\tilde{X} \label{trvec}%
\end{equation}
where we denote $X=(X_{1},\ldots,X_{n})^{T}$ and $\tilde{X}=(\tilde{X}%
_{1},\ldots,\tilde{X}_{n})^{T}$ and where the $n\times n$ matrix $A$ is given above.

All the vector fields $X_{i}$ and $\tilde{X}_{i}$ are naturally tangent to the
corresponding $M_{\alpha,\tilde{\alpha}}$ so that if $\xi_{0}\in$
$M_{\alpha,\tilde{\alpha}}$ then the multi-parameter (simultaneous) solution%
\begin{equation}
\xi=\xi(t_{1},\ldots,t_{n},\xi_{0}) \label{msol}%
\end{equation}
of all equations in (\ref{ham}) starting at $\xi_{0}$ for $t=0$, will always
remain in $M_{\alpha,\tilde{\alpha}}$ and the same is also true for
multi-parameter solutions of (\ref{hamtil}).

The relations (\ref{trXsi})-(\ref{trXi}) can be reformulated in the dual
language, that of reciprocal (multi-time) transformations. The reciprocal
transformation $\tilde{t}_{i}=\tilde{t}_{i}(t_{1},\ldots,t_{n},\xi),$
$i=1,\ldots,n$ given on $M_{\alpha,\tilde{\alpha}}$ by%
\begin{equation}
d\tilde{t}=A^{T}dt \label{recip}%
\end{equation}
where $dt=(dt_{1},\ldots,dt_{n})^{T}$ and $d\tilde{t}=(d\tilde{t}_{1}%
,\ldots,d\tilde{t}_{n})^{T}$, transforms the $k$-parameter solutions
(\ref{msol}) of the system (\ref{ham}) to the $k$-parameter solutions
$\tilde{\xi}=\tilde{\xi}(\tilde{t}_{1},\ldots,\tilde{t}_{n},\xi_{0})$ of the
system (\ref{hamtil}) (with the same initial condition $\xi(0)=\xi_{0}\in
M_{\alpha,\tilde{\alpha}}$) in the sense that for any $\xi_{0}\in
M_{\alpha,\tilde{\alpha}}$ we have%
\[
\tilde{\xi}(\tilde{t}_{1}(t_{1},\ldots,t_{n},\xi_{0}),\ldots,\tilde{t}%
_{n}(t_{1},\ldots,t_{n},\xi_{0}),\xi_{0})=\xi(t_{1},\ldots,t_{n},\xi_{0})
\]
for all values of $t_{i}$ sufficiently close to zero.

The transformation (\ref{recip}) is well defined since the right-hand side of
(\ref{recip}) is an exact differential, as it follows from the above
construction. It means that it is possible (at least locally) to integrate
(\ref{recip}) and obtain an explicit transformation $\tilde{t}_{i}=\tilde
{t}_{i}(t_{1},\ldots,t_{n},\xi)$ that takes multi-time (simultaneous)
solutions of all Hamiltonian systems (\ref{ham}) to multi-time solutions of
all the systems in (\ref{hamtil}).

\section{St\"{a}ckel transform of Lax equations\label{S2}}

In the theorem below, we establish a connection between the Lax pairs of the
systems related by a multi-parameter St\"{a}ckel transform.

\begin{theorem}
\label{T1}Suppose that the Liouville integrable system (\ref{ham}) has the Lax
representation%
\begin{equation}
L_{t_{j}}=\left[  U_{j},L\right]  \text{, \ \ \ }j=1,\ldots,n \label{Lax}%
\end{equation}
where $L=L(\lambda,\xi,\alpha)$ and $U_{j}=U_{j}(\lambda,\xi,\alpha)$ are some
matrices depending on the spectral parameter $\lambda$. Then the Liouville
integrable system (\ref{hamtil}) has the Lax representation%
\begin{equation}
\tilde{L}_{\tilde{t}_{j}}=[\tilde{U}_{j},\tilde{L}]\text{, \ \ \ }j=1,\ldots,n
\label{Laxt}%
\end{equation}
where%
\begin{equation}%
\begin{array}
[c]{c}%
\tilde{L}(\lambda,\xi,\tilde{\alpha})=L(\lambda,\xi,\tilde{h}(\xi
,\tilde{\alpha}))\\
\tilde{U}_{j}(\lambda,\xi,\tilde{\alpha})=%
{\displaystyle\sum\limits_{i=1}^{n}}
(A^{-1})_{ji}(\xi,\tilde{h}(\xi,\tilde{\alpha}))U_{i}(\lambda,\xi,\tilde
{h}(\xi,\tilde{\alpha}))\text{\ }%
\end{array}
\label{LU}%
\end{equation}

\end{theorem}

Thus, in order to obtain the Lax matrix $\tilde{L}(\lambda,\xi,\tilde{\alpha
})$ of the system (\ref{hamtil}) it is enough to replace each $\alpha_{i}$ in
$L(\lambda,\xi,\alpha)$ by the corresponding $\tilde{h}_{s_{i}}(\xi
,\tilde{\alpha})$; the same substitutions are performed in $(A^{-1})_{ji}%
(\xi,\alpha)$ and in $U_{i}(\lambda,\xi,\alpha)$ in the second formula in
(\ref{LU}) in order to obtain $\tilde{U}_{j}(\lambda,\xi,\tilde{\alpha})$.

\begin{proof}
Fix arbitrary values of the parameters $\tilde{\alpha}_{i}$ and choose a point
$\xi\in M$. According to Remark \ref{r} we can then find values of the
parameters $a_{i}$ so that $\xi\in M_{\alpha,\tilde{\alpha}}$ and then
$\alpha_{i}=\tilde{h}_{s_{i}}(\xi,\tilde{\alpha})$ for $i=1,\ldots,k$.
Obviously
\begin{equation}
\partial_{\tilde{t}_{j}}\tilde{h}_{i}=0 \label{1}%
\end{equation}
for all $i,j$ and moreover, due to (\ref{trvec}), we have on $M_{\alpha
,\tilde{\alpha}}$%
\begin{equation}
\frac{\partial}{\partial\tilde{t}_{j}}=\sum_{i=1}^{n}(A^{-1})_{ji}%
\frac{\partial}{\partial t_{i}},\ \ \ j=1,\ldots,n. \label{2a}%
\end{equation}
In consequence, at the chosen (and thus arbitrary) $\xi\in M$
\begin{align*}
\tilde{L}_{\tilde{t}_{j}}(\lambda,\xi,\tilde{\alpha})  &  =L_{\tilde{t}_{j}%
}(\lambda,\xi,\tilde{h}(\xi,\tilde{\alpha}))\overset{(\ref{1})}{=}L_{\tilde
{t}_{j}}(\lambda,\xi,\alpha)\overset{(\ref{2a})}{=}%
{\displaystyle\sum\limits_{i=1}^{n}}
A^{-1}(\xi,\alpha){}_{ji}L_{t_{i}}(\lambda,\xi,\alpha)\overset{(\ref{Lax})}%
{=}\\
&  =%
{\displaystyle\sum\limits_{i=1}^{n}}
A^{-1}(\xi,\alpha)_{ji}\left[  U_{i}(\lambda,\xi,\alpha),L(\lambda,\xi
,\alpha)\right]  =\left[  \tilde{U}_{j}(\lambda,\xi,\tilde{\alpha}),\tilde
{L}(\lambda,\xi,\tilde{\alpha})\right]
\end{align*}
Let us make two comments on the above theorem. Firstly, the Lax pairs
(\ref{Lax}) and (\ref{Laxt}) are understood as differential-algebraic
consequences of the systems (\ref{ham}) and (\ref{hamtil}) respectively, i.e.
we do not require that the Lax pairs (\ref{Lax}) and (\ref{Laxt}) actually
reconstruct the systems themselves (see also Remark \ref{wazny}). Further,
this theorem is a global result, not just restricted to some submanifold
$M_{\alpha,\tilde{\alpha}}$.
\end{proof}

\section{Hamiltonian systems from the seed class and their Lax
representations\label{S3}}

Let us now consider separable systems generated by separation curves (spectral
curves) in the form
\begin{equation}
\sigma(\lambda)+\sum_{j=1}^{n}H_{j}\lambda^{n-j}=\frac{1}{2}f(\lambda)\mu^{2}.
\label{B}%
\end{equation}
Solving the system of $n$ copies of (\ref{B}), with $\lambda_{i}$ and $\mu
_{i}$ substituted for $\lambda$ and $\mu$, $i=1,\ldots,n$, with respect to
$H_{j}$ we obtain $n$ separable (and thus Liouville integrable) Hamiltonians
$H_{j}(\lambda,\mu)$ and $n$ related vector fields $X_{j}(\lambda,\mu),$
\begin{equation}
H_{j}(\lambda,\mu)=E_{j}(\lambda,\mu)+V_{j}^{(\sigma)}(\lambda),\ \ X_{j}%
(\lambda,\mu)=\pi dH_{j}(\lambda,\mu)\text{, \ \ }j=1,...,n \label{Bh}%
\end{equation}
on the Poisson manifold $(M,\pi),$ where $E_{j}(\lambda,\mu)$ represent
geodesic part of the Hamiltonians and $V_{j}^{(\sigma)}(\lambda)$ represent
the potential part. Throughout the article $(\lambda,\mu)=(\lambda_{1}%
,\ldots,\lambda_{n},\mu_{1},\ldots,\mu_{n})$ denotes Darboux (canonical)
coordinates on $(M,\pi)$ which are also separation coordinates for all $H_{j}$
\cite{blasz2005}. The functions $V_{j}^{(\sigma)}$ are linear combinations of
so called basic separable potentials $V_{j}^{(\beta)}$, generated by monomials
$\sigma(\lambda)=\lambda^{\beta}$, $\beta\in\mathbb{Z}$. The potentials
$V_{j}^{(\beta)}$ can be obtained by solving $n$ copies of the equation%
\begin{equation}
\lambda^{\beta}+\sum_{j=1}^{n}V_{j}^{(\beta)}\lambda^{n-j}=0,\ \ \ \beta
\in\mathbb{Z} \label{V}%
\end{equation}
Explicitly the functions $V_{j}^{(\beta)}$ can be calculated from the formula
\cite{blasz2011}
\begin{equation}
V^{(\beta)}=R^{\beta}V^{(0)},\ \ \ \ V^{(\beta)}=(V_{1}^{(\beta)}%
,...,V_{n}^{(\beta)})^{T}, \label{6}%
\end{equation}
where
\begin{equation}
R=\left(
\begin{array}
[c]{cccc}%
-\rho_{1}(\lambda) & 1 & 0 & 0\\
\vdots & 0 & \ddots & 0\\
\vdots & 0 & 0 & 1\\
-\rho_{n}(\lambda) & 0 & 0 & 0
\end{array}
\right)  ,\ \ \ \ \ V^{(0)}=(0,0,...,-1)^{T}, \label{6a}%
\end{equation}
$\rho_{i}(\lambda)=(-1)^{i}\sigma_{i}(\lambda)$ and $\sigma_{i}(\lambda)$ are
elementary symmetric polynomials. Notice that for $\beta=0,...,n-1$%
\begin{equation}
V_{k}^{(\beta)}=-\delta_{k,n-\beta}. \label{7a}%
\end{equation}

It has been proved in \cite{blasz2018} that each system (\ref{Bh}) has a
family of Lax representations%
\begin{equation}
\frac{d}{dt_{k}}L(\lambda)=[U_{k}(\lambda),L(\lambda)],\text{ \ \ }%
k=1,\ldots,n, \label{25}%
\end{equation}
parametrized by arbitrary nonvanishing smooth functions $g(\lambda)$. The Lax
matrix in the Darboux variables $(\lambda,\mu)$, has the form
\cite{blasz2018}
\begin{equation}
L(\lambda)=%
\begin{pmatrix}
v(\lambda) & u(\lambda)\\
w(\lambda) & -v(\lambda)
\end{pmatrix}
, \label{14}%
\end{equation}
while the auxiliary matrices $U_{k}(\lambda)$ in (\ref{25}) are of the form
\begin{equation}
U_{k}(\lambda)=\left[  \frac{B_{k}(\lambda)}{u(\lambda)}\right]  _{+},\quad
B_{k}(\lambda)=\frac{1}{2}\frac{f(\lambda)}{g(\lambda)}\left[  \frac
{u(\lambda)}{\lambda^{n-k+1}}\right]  _{+}L(\lambda) \label{125a}%
\end{equation}
where
\begin{equation}
u(\lambda)\equiv\prod\limits_{k=1}^{n}(\lambda-\lambda_{k})=\sum_{k=0}^{n}%
\rho_{k}\lambda^{n-k},\quad\rho_{0}\equiv1 \label{15a}%
\end{equation}
and
\begin{equation}
v(\lambda)=\sum_{i=1}^{n}g(\lambda_{i})\mu_{i}\prod\limits_{k\neq i}%
\frac{\lambda-\lambda_{k}}{\lambda_{i}-\lambda_{k}}=-\sum_{k=0}^{n-1}\left[
\sum_{i=1}^{n}\frac{\partial\rho_{n-k}}{\partial\lambda_{i}}\frac
{g(\lambda_{i})\mu_{i}}{\Delta_{i}}\right]  \lambda^{k} \label{15}%
\end{equation}
(notice that $u(\lambda_{i})=0$ and $v(\lambda_{i})=g(\lambda_{i})\mu_{i}$)
while
\begin{equation}
w(\lambda)=-2\frac{g^{2}(\lambda)}{f(\lambda)}\left[  \frac{F(\lambda
,v(\lambda)/g(\lambda))}{u(\lambda)}\right]  _{+}, \label{21}%
\end{equation}
where $F(x,y)=\frac{1}{2}f(x)y^{2}-\sigma(x)$. The symbol $[\frac{{}}{{}}%
]_{+}$ denotes a polynomial part (Laurent polynomial part) of the quotient,
i.e. if $P(\lambda)$ is a polynomial or a Laurent polynomial and if
$Q(\lambda)$ is a polynomial then:%
\begin{equation}
\frac{P(\lambda)}{Q(\lambda)}=\left[  \frac{P(\lambda)}{Q(\lambda)}\right]
_{+}+\frac{R(\lambda)}{Q(\lambda)}\text{ } \label{D}%
\end{equation}
where $R(\lambda)$ is a reminder of the quotient, so $\deg R<\deg Q$ (see
\cite{blasz2018} for details). In particular, for positive basic separable
potentials $V_{j}^{(n+s)}$, $s\in\mathbb{N}$ we have
\begin{equation}
\left[  \frac{\lambda^{n+s}}{u(\lambda)}\right]  _{+}=-\sum_{r=0}^{s}%
V_{1}^{(n+r-1)}\lambda^{s-r}=-\sum_{r=0}^{s}V_{1}^{(n+s-r-1)}\lambda^{r}
\label{23b}%
\end{equation}
while for basic negative separable potentials $V_{j}^{(-s)}$, $s\in\mathbb{N}$%
\begin{equation}
\left[  \frac{\lambda^{-s}}{u(\lambda)}\right]  _{+}=\sum_{r=1}^{s}%
V_{1}^{(-r)}\lambda^{-s+r-1}=\sum_{r=1}^{s}V_{1}^{(-s+r-1)}\lambda^{-r}.
\label{24b}%
\end{equation}
Notice also that the function $w(\lambda)$ in (\ref{21}) splits into kinetic
part $w_{E}(\lambda)$ and potential part $w_{V}(\lambda)$ respectively:
\begin{equation}
w(\lambda)=w_{E}(\lambda)+w_{V}(\lambda)=-\frac{g^{2}(\lambda)}{f(\lambda
)}\left[  \frac{f(\lambda)v^{2}(\lambda)}{u(\lambda)g^{2}(\lambda)}\right]
_{+}+2\frac{g^{2}(\lambda)}{f(\lambda)}\left[  \frac{\sigma(\lambda
)}{u(\lambda)}\right]  _{+}. \label{20}%
\end{equation}

\begin{remark}
\label{wazny}The Lax matrix $L(\lambda)$ (\ref{14}) reconstructs the
separation curve (\ref{B}) in the sense that \cite{blasz2018}
\begin{equation}
0=\det\left[  L(\lambda)-g(\lambda)\mu I\right]  =-2\frac{g^{2}(\lambda
)}{f(\lambda)}\left(  \sigma(\lambda)+\sum_{j=1}^{n}H_{j}\lambda^{n-j}%
-\frac{1}{2}f(\lambda)\mu^{2}\right)  . \label{22b}%
\end{equation}
The Lax matrices for different choices of $g(\lambda)$ are not equivalent.
\end{remark}

The Lax pairs (\ref{25}), although given here in the separation coordinates
$(\lambda,\mu)$, are invariant with respect to any change of coordinates on
the manifold $M$. In particular, we will use so called Viet\'{e} coordinates
defined as
\begin{equation}
q_{i}=(-1)^{i}\sigma_{i}(\lambda)\text{, \ \ }p_{i}=-\sum_{k=1}^{n}%
\frac{\lambda_{k}^{n-i}\mu_{k}}{\Delta_{k}},\text{ \ \ }\Delta_{i}%
=\prod\limits_{k\neq i}(\lambda_{i}-\lambda_{k})\text{, \ \ }%
\ i=1,...,n.\ \label{Viete}%
\end{equation}
In these coordinates \cite{blasz2018}
\begin{equation}
u(\lambda;q)=\sum_{k=0}^{n}q_{k}\lambda^{n-k},\quad q_{0}\equiv1 \label{V1}%
\end{equation}
and
\begin{equation}
v(\lambda;q,p)=\sum_{k=1}^{n}\left[  \sum_{s=0}^{k-1}q_{s}\left(  \sum
_{j=1}^{n}V_{j}^{(r+k-s-1)}p_{j}\right)  \right]  \lambda^{n-k},\ \ \ \ \ r\in
\mathbb{Z} \label{V2}%
\end{equation}

\section{St\"{a}ckel transform of the seed class\label{S4}}

In the previous sections we have discussed systems from the seed class,
generated by separation curves (\ref{B}), and their Lax pairs (\ref{25}). In
this section we will demonstrate, using this knowledge, how to generate
systems from $\gamma$-classes (\ref{Si}), defined by the separation curves of
the form%
\begin{equation}
\sigma(\lambda)+%
{\displaystyle\sum\limits_{j=1}^{n}}
\tilde{H}_{j}\lambda^{\gamma_{j}}=\frac{1}{2}f(\lambda)\mu^{2}\text{, }
\label{S}%
\end{equation}
where $\gamma_{i}\in\mathbb{N}$ and $\gamma_{1}>\gamma_{2}>\cdots>\gamma
_{n}=0$. Actually, our goal is to demonstrate how to construct the St\"{a}ckel
systems of a given $\gamma$-class (\ref{S}) by applying the multi-parameter
St\"{a}ckel transform (\ref{alphy})-(\ref{reszta}) to an appropriate system
from the seed class (\ref{B}).

In order to be able to relate the systems from classes (\ref{B}) and (\ref{S})
by a St\"{a}ckel transform, we need to extend both of them to appropriate
multi-parameter systems.

\begin{theorem}
\label{T2}Assume that $\gamma_{1}>\gamma_{2}>\cdots>\gamma_{k}>n-1$ and
$S=\{s_{1},\dots,s_{k}\}\subset\{1,\dots,n\}\ $\ with $s_{1}<\cdots<s_{k}$.
Then, the St\"{a}ckel transform (\ref{alphy})-(\ref{reszta}) transforms the
Hamiltonians
\begin{equation}
h_{j}(\lambda,\mu,\alpha)=H_{j}(\lambda,\mu)+%
{\displaystyle\sum\limits_{i=1}^{k}}
\alpha_{i}V_{j}^{(\gamma_{i})}, \label{BRh}%
\end{equation}
defined by the separation curve from the seed class
\begin{equation}
\sigma(\lambda)+%
{\displaystyle\sum\limits_{j=1}^{k}}
\alpha_{j}\lambda^{\gamma_{j}}+%
{\displaystyle\sum\limits_{j=1}^{n}}
h_{j}\lambda^{n-j}=\frac{1}{2}f(\lambda)\mu^{2}, \label{BR}%
\end{equation}
to the Hamiltonians
\begin{equation}
\tilde{h}_{j}(\lambda,\mu,\alpha)=\tilde{H}_{j}(\lambda,\mu)+%
{\displaystyle\sum\limits_{i=1}^{k}}
\tilde{\alpha}_{i}\tilde{V}_{j}^{(n-s_{i})}. \label{SRh}%
\end{equation}
defined by the separation curve from the $\gamma$-class $\gamma=\left\{
\gamma_{1},\ldots,\gamma_{k}\right\}  \cup\left\{  n-i:i\notin S\right\}  $
\begin{equation}
\sigma(\lambda)+%
{\displaystyle\sum\limits_{j=1}^{n}}
\tilde{h}_{j}\lambda^{\gamma_{j}}+\sum_{j=1}^{k}\tilde{\alpha}_{j}%
\lambda^{n-s_{j}}=\frac{1}{2}f(\lambda)\mu^{2}, \label{SR}%
\end{equation}
Moreover, the explicit transform between Hamiltonians from both classes takes
the form%
\begin{equation}
\tilde{h}=-A_{\gamma}^{-1}H+A_{\gamma}^{-1}\tilde{\alpha}, \label{st}%
\end{equation}
where
\[
h=H+A_{\gamma}\alpha
\]
and \ $h=(h_{1},\ldots,h_{n})^{T}$, $H=(H_{1},\ldots,H_{n})^{T}$,
$\alpha=(\alpha_{1},...,\alpha_{k},0,...,0)^{T}$ and likewise for $\tilde{h}$,
$\tilde{H}$, $\tilde{\alpha}$. The $n\times n$ matrix $A_{\gamma}$ is given
by
\begin{equation}
\left(  A_{\gamma}\right)  _{ij}=V_{i}^{(\gamma_{j})} \label{Ag}%
\end{equation}
\ and in particular we obtain the explicit map between $H$ and $\tilde{H}$:
\begin{equation}
\tilde{H}=-A_{\gamma}^{-1}H. \label{SH}%
\end{equation}

\end{theorem}

This theorem follows from Theorem 3 in\ \cite{blasz2012}. A careful inspection
of this theorem reveals that the enumeration of Hamiltonians $\tilde{h}_{i}$
in this theorem has been changed in comparison with the general construction
presented in Section \ref{S1}. We still have that $h_{s_{i}}=\tilde{\alpha
}_{i}$ for $i=1,\ldots,k$ but now $\tilde{h}_{i}=\alpha_{i}$ for
$i=1,\ldots,k$ (and not $\tilde{h}_{_{s_{i}}}=\alpha_{i}$ as in the general
construction) while the remaining transformed Hamiltonians $\tilde{h}_{i}$
(for $i=k+1,\ldots,n$) are obtained by substituting, in the consecutive
$h_{j}$ for $j\notin S$, all $\alpha_{i}$, $i=1,\ldots,k$, with the
corresponding $\tilde{h}_{i}$, as the general idea of St\"{a}ckel transform
stipulates. This is done in order to obtain a convenient enumeration of
$\tilde{h}_{i}$ in the transformed system (\ref{SR}).

Notice that due to (\ref{BRh}) the matrix $A_{\gamma}$ is simply the matrix
$A$ (\ref{A}) written in the particular settings of this theorem. Notice also
that (\ref{SH}) is valid on the whole $M$, in contrast with the relations
(\ref{trdh}) that are valid only on $M_{\alpha,\tilde{\alpha}}$. However, by
explanations in Section \ref{S1}, the solutions of (\ref{BRh}) and (\ref{SRh})
are related only on the appropriate submanifolds $M_{a,\tilde{\alpha}}$. Thus,
although the St\"{a}ckel transform (\ref{SH}) transforms the parameter-free
Liouville-integrable system (\ref{B}) into another parameter-free Liouville
integrable system (\ref{S}), the solutions of these systems are not globally
related by any reciprocal transformation.

Let us demonstrate the whole procedure on two examples, both involving
one-parameter St\"{a}ckel transform. We restrict ourselves to one-parameter
examples as the examples involving two parameters lead to large and
complicated expressions not very suitable to be presented in a printed form.

\begin{example}
\label{E0}As a first example, consider the H\'{e}non-Heiles system given by
the separation curve%
\begin{equation}
H_{1}\lambda+H_{2}=\frac{1}{2}\lambda\mu^{2}+\lambda^{4} \label{hh}%
\end{equation}
(so that $n=2$, $\sigma(\lambda)=-\lambda^{4}$ and $f(\lambda)=\lambda$) and
its one-parameter (so that $k=1$) St\"{a}ckel transform ("one-hole
deformation" \cite{blasz2005})) with $\gamma_{1}=2$ and with respect to the
first Hamiltonian, i.e. $S=\left\{  1\right\}  $. Thus $\gamma=\{2,0\}$ and
the extended H\'{e}non-Heiles system is generated by separation curve
\begin{equation}
\alpha\lambda^{2}+h_{1}\lambda+h_{2}=\frac{1}{2}\lambda\mu^{2}+\lambda^{4}
\label{hhe}%
\end{equation}
The St\"{a}ckel transform (\ref{st}) yields the $\gamma$-system generated by%
\begin{equation}
\tilde{h}_{1}\lambda^{2}+\tilde{\alpha}\lambda+\widetilde{h}_{2}=\frac{1}%
{2}\lambda\mu^{2}+\lambda^{4} \label{shhe}%
\end{equation}
Setting $\tilde{\alpha}=0$ we obtain the system generated by
\begin{equation}
\tilde{H}_{1}\lambda^{2}+\tilde{H}_{2}=\frac{1}{2}\lambda\mu^{2}+\lambda^{4}
\label{shh}%
\end{equation}
The matrix $A_{\gamma}$ given by (\ref{Ag}) is
\[
A_{\gamma}=\left(
\begin{array}
[c]{cc}%
V_{1}^{(2)} & 0\\
V_{2}^{(2)} & -1
\end{array}
\right)
\]
and thus the parameter-free St\"{a}ckel transform (\ref{SH}) between both
systems is%
\[
\tilde{H}_{1}=-\frac{1}{V_{1}^{(2)}}H_{1},\ \ \ \ \tilde{H}_{2}=H_{2}%
-\frac{V_{2}^{(2)}}{V_{1}^{(2)}}H_{1}.
\]
We will illustrate the structure of the above objects in the flat orthogonal
coordinates $(x,y)$ of the system. They are given by%
\begin{equation}
x_{1}=\lambda_{1}+\lambda_{2}\text{, \ \ }x_{2}=2\sqrt{-\lambda_{1}\lambda
_{2}} \label{xy}%
\end{equation}
with the conjugate momenta given by%
\begin{equation}
y_{1}=\frac{\lambda_{1}\mu_{1}}{\lambda_{1}-\lambda_{2}}+\frac{\lambda_{2}%
\mu_{2}}{\lambda_{2}-\lambda_{1}}\text{, \ \ }y_{2}=\sqrt{-\lambda_{1}%
\lambda_{2}}\left(  \frac{\mu_{1}}{\lambda_{1}-\lambda_{2}}+\frac{\mu_{2}%
}{\lambda_{2}-\lambda_{1}}\right)  \label{ygreki}%
\end{equation}
In the flat coordinates the Hamiltonians $H_{i}$ attain the form%
\begin{align}
H_{1}  &  =\frac{1}{2}y_{1}^{2}+\frac{1}{2}y_{2}^{2}+x_{1}^{3}+\frac{1}%
{2}x_{1}x_{2}^{2}\nonumber\\
H_{2}  &  =\frac{1}{2}x_{2}y_{1}y_{2}-\frac{1}{2}x_{1}y_{2}^{2}+\frac{1}%
{16}x_{2}^{4}+\frac{1}{4}x_{1}^{2}x_{2}^{2} \label{hh1}%
\end{align}
while their St\"{a}ckel transform becomes%
\begin{align}
\tilde{H}_{1}  &  =\frac{1}{2}\frac{1}{x_{1}}y_{1}^{2}+\frac{1}{2}\frac
{1}{x_{1}}y_{2}^{2}+x_{1}^{2}+\frac{1}{2}x_{2}^{2}\nonumber\\
\ \tilde{H}_{2}  &  =-\frac{1}{8}\frac{x_{2}^{2}}{x_{1}}y_{1}^{2}+\frac{1}%
{2}x_{1}^{2}y_{1}y_{2}-\frac{1}{8}\frac{x_{2}^{2}}{x_{1}}y_{2}^{2}-\frac
{1}{16}x_{2}^{4} \label{shh1}%
\end{align}

\end{example}

\begin{example}
\label{E1}Consider now the one-parameter (again $k=1$) St\"{a}ckel transform
of the system defined by separation curve (\ref{B})
\begin{equation}
\lambda^{5}+H_{1}\lambda^{2}+H_{2}\lambda+H_{3}=\frac{1}{2}\mu^{2} \label{e1}%
\end{equation}
(so that $n=3$, $\sigma(\lambda)=\lambda^{5}$ and $f(\lambda)=1$), with
$\gamma_{1}=3$ and with respect to the second Hamiltonian, i.e. $S=\{2\}$.
Thus $\gamma=\left\{  3,2,0\right\}  $ so the extended system is defined by
separation curve (\ref{BR})
\begin{equation}
\lambda^{5}+\alpha\lambda^{3}+h_{1}\lambda^{2}+h_{2}\lambda+h_{3}=\frac{1}%
{2}\mu^{2} \label{r1}%
\end{equation}
and the St\"{a}ckel transform (\ref{st}) yields the $\gamma$-system defined by
(\ref{SR})%
\begin{equation}
\lambda^{5}+\tilde{h}_{1}\lambda^{3}+\tilde{h}_{2}\lambda^{2}+\tilde{\alpha
}\lambda+\tilde{h}_{3}=\frac{1}{2}\mu^{2} \label{r2}%
\end{equation}
Setting $\tilde{\alpha}=0$ we obtain a new system defined by separation curve
\begin{equation}
\lambda^{5}+\tilde{H}_{1}\lambda^{3}+\tilde{H}_{2}\lambda^{2}+\tilde{H}%
_{3}=\frac{1}{2}\mu^{2} \label{e2}%
\end{equation}
As
\[
A_{\gamma}=\left(
\begin{array}
[c]{ccc}%
V_{1}^{(3)} & -1 & 0\\
V_{2}^{(3)} & 0 & 0\\
V_{3}^{(3)} & 0 & -1
\end{array}
\right)
\]
so the parameter-free St\"{a}ckel transform (\ref{SH}) between both systems
attains the form:
\[
\tilde{H}_{1}=-\frac{1}{V_{2}^{(3)}}H_{2},\ \ \ \ \tilde{H}_{2}=H_{1}%
-\frac{V_{1}^{(3)}}{V_{2}^{(3)}}H_{2},\ \ \ \ \ \tilde{H}_{3}=H_{3}%
-\frac{V_{3}^{(3)}}{V_{2}^{(3)}}H_{2}.
\]
As in the previous example, we will explicitly illustrate the structure of
both systems in another coordinates. This time we make a point transformation
to Viet\'{e} coordinates (\ref{Viete}). In these coordinates all the
Hamiltonians $H_{i}$ are polynomials \cite{blasz2007}. Explicitly
\begin{align}
H_{1}  &  =\frac{1}{2}{p_{{2}}^{2}+}\,q_{{1}}p_{{2}}p_{{3}}+\,p_{{1}}%
p_{3}+\frac{1}{2}q_{{2}}{p_{{3}}^{2}}+{q_{{1}}^{3}}-2\,q_{{1}}q_{{2}}+q_{{3}%
}\nonumber\\
H_{2}  &  =q_{{1}}\,p_{{1}}p_{{3}}+q_{{1}}\,{p_{{2}}^{2}}+p_{{1}}\,p_{{2}%
}+\frac{1}{2}\left(  \,\,q_{{1}}q_{{2}}-q_{{3}}\right)  {p_{{3}}^{2}%
}+\,{q_{{1}}^{2}}p_{{2}}p_{{3}}+{q_{{1}}^{2}}q_{{2}}-q_{{1}}q_{{3}}-{q_{{2}%
}^{2}}\label{s1}\\
H_{3}  &  =\frac{1}{2}{p_{{1}}^{2}+}\,q_{{2}}p_{{1}}p_{{3}}+\frac{1}{2}%
{q_{{1}}^{2}p_{{2}}^{2}}+2q_{{1}}p_{{1}}p_{{2}}+\frac{1}{2}\left(  \,{q_{{2}%
}^{2}}-q_{{1}}q_{{3}}\right)  {p_{{3}}^{2}}+\,\left(  q_{{1}}q_{{2}}-q_{{3}%
}\right)  p_{{2}}p_{{3}}+{q_{{1}}^{2}}q_{{3}}-q_{{2}}q_{{3}}\nonumber
\end{align}
while%
\begin{align}
\tilde{H}_{1}  &  =-\frac{1}{q_{{2}}}{p}_{1}p_{2}-{\frac{q_{{1}}}{q_{{2}}}%
}p_{1}p_{3}-\,{\frac{q_{{1}}}{q_{{2}}}}p_{2}^{2}-{\frac{{q_{{1}}^{2}}}{q_{{2}%
}}}p_{{2}}p_{{3}}+\frac{1}{2}\left(  {\frac{q_{{3}}}{q_{{2}}}}-q_{{1}}\right)
{p_{{3}}^{2}}-{q_{{1}}^{2}}+{\frac{q_{{1}}q_{{3}}}{q_{{2}}}}+q_{{2}%
}\nonumber\\
\tilde{H}_{2}  &  =-\,{\frac{q_{{1}}}{q_{{2}}}}p_{{1}}p_{{2}}+\left(
1-\,{\frac{{q_{{1}}^{2}}}{q_{{2}}}}\right)  p_{{1}}p_{{3}}+\left(  \frac{1}%
{2}-{\frac{{q_{{1}}^{2}}}{q_{{2}}}}\right)  {p_{{2}}^{2}}+\left(  \,q_{{1}%
}-{\frac{{q_{{1}}^{3}}}{q_{{2}}}}\right)  p_{{2}}p_{{3}}\nonumber\\
&  +\frac{1}{2}\left(  {\frac{q_{{3}}q_{{1}}}{q_{{2}}}}+q_{{2}}-{q_{{1}}^{2}%
}\right)  {p_{{3}}^{2}}+{\frac{q_{{3}}{q_{{1}}^{2}}}{q_{{2}}}}-q_{{1}}q_{{2}%
}+q_{{3}}\label{s2}\\
\tilde{H}_{3}  &  =\frac{1}{2}{p_{{1}}^{2}}+\left(  q_{{1}}-{\frac{q_{{3}}%
}{q_{{2}}}}\right)  p_{{1}}p_{{2}}+\left(  \,q_{{2}}-{\frac{q_{{1}}q_{{3}}%
}{q_{{2}}}}\right)  p_{{1}}p_{{3}}+\left(  \frac{1}{2}{q_{{1}}^{2}}%
-{\frac{q_{{1}}q_{{3}}}{q_{{2}}}}\right)  {p_{{2}}^{2}}\nonumber\\
&  +\left(  \,q_{{1}}q_{{2}}-\,q_{{3}}-\,{\frac{{q_{{1}}^{2}}q_{{3}}}{q_{{2}}%
}}\right)  p_{{2}}p_{{3}}+\left(  \frac{1}{2}{q_{{2}}^{2}}+\frac{1}{2}%
{\frac{{q_{{3}}^{2}}}{q_{{2}}}}-q_{{1}}\,q_{{3}}\right)  {p_{{3}}^{2}}%
+{\frac{q_{{1}}{q_{{3}}^{2}}}{q_{{2}}}.}\nonumber
\end{align}

\end{example}

\section{Lax representation of $\gamma$-classes\label{S5}}

In this section we apply the results from Section \ref{S2} and Section
\ref{S4}\ in order to construct the Lax pairs for the St\"{a}ckel system
(\ref{S}) from a given $\gamma$-class. In order to do this we start from the
Lax pairs $(L(\lambda,\alpha)$, $U_{j}(\lambda,\alpha))$ for the extended
systems from the seed class (\ref{BR}), where%
\begin{equation}
L(\lambda,\alpha)=\left(
\begin{array}
[c]{cc}%
v(\lambda) & u(\lambda)\\
w(\lambda,\alpha) & -v(\lambda)
\end{array}
\right)  \label{La}%
\end{equation}
with
\[
w(\lambda,\alpha)=w(\lambda)+2\frac{g^{2}(\lambda)}{f(\lambda)}\sum
\nolimits_{j=1}^{k}\alpha_{j}\left[  \frac{\lambda^{\gamma_{j}}}{u(\lambda
)}\right]  _{+}%
\]
and $U_{j}(\lambda,\alpha)$ are given by (\ref{125a})\ with $L(\lambda)$
replaced by $L(\lambda,\alpha)$. From Theorem \ref{T1} we obtain the following corollary.

\begin{corollary}
For any smooth nonvanishing function $g(\lambda)$ the matrices $\left(
\tilde{L}(\lambda,\tilde{\alpha}),\tilde{U}_{j}(\lambda,\tilde{\alpha
})\right)  $ given by%
\begin{equation}
\tilde{L}(\lambda,\tilde{\alpha})=\left(
\begin{array}
[c]{cc}%
v(\lambda) & u(\lambda)\\
\tilde{w}(\lambda,\tilde{\alpha}) & -v(\lambda)
\end{array}
\right)  \label{Lat}%
\end{equation}
where%
\[
\tilde{w}(\lambda,\tilde{\alpha})=w(\lambda)+2\frac{g^{2}(\lambda)}%
{f(\lambda)}\sum\nolimits_{j=1}^{k}\tilde{h}_{j}(\xi,\tilde{\alpha})\left[
\frac{\lambda^{\gamma_{j}}}{u(\lambda)}\right]  _{+}%
\]
and with
\[
\tilde{U}_{j}(\lambda,\tilde{\alpha})=%
{\displaystyle\sum\limits_{i=1}^{n}}
\left(  A^{-1}\right)  _{ji}(\xi)U_{i}(\lambda,\tilde{h}(\xi,\tilde{\alpha
})).
\]
constitute the Lax pairs for the extended systems from the $\gamma$-class
(\ref{SR}).
\end{corollary}

The matrices $U_{i}(\lambda,\tilde{h}(\xi,\tilde{\alpha}))$ can effectively be
calculated by the formulas (\ref{125a})\ with $L(\lambda)$ replaced by
$\tilde{L}(\lambda,\tilde{\alpha})$. Notice that the matrix~$A$ in the above
formula does not depend on $\alpha$ (it still does depend on $\xi$). Finally,
the Lax pairs $\left(  \tilde{L}(\lambda),\tilde{U}_{j}(\lambda)\right)  $ for
the system (\ref{SR}) are obtained by letting $\tilde{\alpha}_{i}=0$ in
$\left(  \tilde{L}(\lambda,\tilde{\alpha}),\tilde{U}_{j}(\lambda,\tilde
{\alpha})\right)  $.

\begin{example}
\label{E1.5}(Example \ref{E0} continued) The H\'{e}non-Heiles system
(\ref{hh}) has the Lax pairs given by (\ref{14}-\ref{20}). Then for the
extended system (\ref{hhe}) we get
\[
L(\lambda,\alpha)=\left(
\begin{array}
[c]{cc}%
v(\lambda) & u(\lambda)\\
& \\
w(\lambda)+2\alpha\frac{g^{2}(\lambda)}{\lambda}\left[  \frac{\lambda^{2}%
}{u(\lambda)}\right]  _{+} & -v(\lambda)
\end{array}
\right)
\]
(and $U_{j}$ as given by (\ref{125a})) and thus, for the extended $\gamma
$-system (\ref{shhe}) the Lax matrices are%
\[
\tilde{L}(\lambda,\tilde{\alpha})=\left(
\begin{array}
[c]{cc}%
v(\lambda) & u(\lambda)\\
& \\
w(\lambda)+2\tilde{h}_{1}\frac{g^{2}(\lambda)}{\lambda}\left[  \frac
{\lambda^{2}}{u(\lambda)}\right]  _{+} & -v(\lambda)
\end{array}
\right)
\]
and%
\[
\tilde{U}_{1}(\lambda)=-\frac{1}{V_{1}^{(2)}}U_{1}(\lambda,\alpha=\tilde
{h}_{1}),\ \ \ \ \tilde{U}_{2}(\lambda)=U_{2}(\lambda,\alpha=\tilde{h}%
_{1})-\frac{V_{2}^{(2)}}{V_{1}^{(2)}}U_{1}(\lambda,\alpha=\tilde{h}_{1})
\]
so that the Lax matrices for the transformed system (\ref{shh}) (or
(\ref{shh1})) are
\[
\tilde{L}(\lambda)=\left(
\begin{array}
[c]{cc}%
v(\lambda) & u(\lambda)\\
& \\
w(\lambda)+2\tilde{H}_{1}\frac{g^{2}(\lambda)}{\lambda}\left[  \frac
{\lambda^{2}}{u(\lambda)}\right]  _{+} & -v(\lambda)
\end{array}
\right)
\]
and
\[
\tilde{U}_{1}(\lambda)=-\frac{1}{V_{1}^{(2)}}U_{1}(\lambda,\alpha=\tilde
{H}_{1}),\ \ \ \ \tilde{U}_{2}(\lambda)=U_{2}(\lambda,\alpha=\tilde{H}%
_{1})-\frac{V_{2}^{(2)}}{V_{1}^{(2)}}U_{1}(\lambda,\alpha=\tilde{H}_{1}).
\]
As in Example \ref{E0}, we present the explicit form of these matrices for a
particular function $g(\lambda)$ and in the flat coordinates $(x,y)$, given by
(\ref{xy})-(\ref{ygreki}). Thus, for the H\'{e}non-Heiles system (\ref{hh})
(or equivalently (\ref{hh1})), the Lax matrix $L(\lambda)$ for $g(\lambda
)=1$\ takes the form \cite{blasz2018}
\[
L(\lambda)=%
\begin{pmatrix}
\frac{2y_{2}}{x_{2}}\lambda+y_{1}-\frac{2x_{1}y_{2}}{x_{2}} & \lambda
^{2}-x_{1}\lambda-\frac{1}{4}x_{2}^{2}\\
& \\
-2\lambda-\left(  \frac{4y_{2}^{2}}{x_{2}^{2}}+2x_{1}\right)  +\left(
\frac{4x_{1}y_{2}^{2}}{x_{2}^{2}}-\frac{4y_{1}y_{2}}{x_{2}}-2x_{1}^{2}%
-\frac{1}{2}x_{2}^{2}\right)  \lambda^{-1} & -\frac{2y_{2}}{x_{2}}%
\lambda-y_{1}+\frac{2x_{1}y_{2}}{x_{2}}%
\end{pmatrix}
\]
while
\[
U_{1}(\lambda)=%
\begin{pmatrix}
\frac{y_{2}}{x_{2}} & \frac{1}{2}\lambda\\
& \\
-1 & -\frac{y_{2}}{x_{2}}%
\end{pmatrix}
,\quad U_{2}(\lambda)=%
\begin{pmatrix}
\frac{y_{2}}{x_{2}}\lambda-\frac{x_{1}y_{2}}{x_{2}}+\frac{1}{2}y_{1} &
\frac{1}{2}\lambda^{2}-\frac{1}{2}x_{1}\lambda\\
& \\
-\lambda-\frac{2y_{2}^{2}}{x_{2}^{2}}-x_{1} & -\frac{y_{2}}{x_{2}}%
\lambda+\frac{x_{1}y_{2}}{x_{2}}-\frac{1}{2}y_{1}%
\end{pmatrix}
\text{. }%
\]
The St\"{a}ckel transform (\ref{LU})\ of these Lax pairs yields
\[
\tilde{L}(\lambda)=\left(
\begin{array}
[c]{cc}%
\frac{2y_{2}}{x_{2}}\lambda-\left(  \frac{2x_{1}y_{2}}{x_{2}}-y_{1}\right)  &
\lambda^{2}-x_{1}\lambda-\frac{1}{4}x_{2}^{2}\\
& \\
-2\lambda-\left(  2x_{1}+\frac{4y_{2}^{2}}{x_{2}^{2}}\right)  +\left(
\frac{1}{2}x_{2}^{2}+\frac{4x_{1}y_{2}^{2}}{x_{2}^{2}}-\frac{4y_{1}y_{2}%
}{x_{2}}+\frac{y_{1}^{2}}{x_{1}}+\frac{y_{2}^{2}}{x_{1}}\right)  \lambda^{-1}
& -\frac{2y_{2}}{x_{2}}\lambda+\left(  \frac{2x_{1}y_{2}}{x_{2}}-y_{1}\right)
\end{array}
\right)
\]
and
\[
\tilde{U}_{1}(\lambda)=\left(
\begin{array}
[c]{cc}%
\frac{y_{2}}{x_{1}x_{2}} & \frac{1}{2}\frac{1}{x_{1}}\lambda\\
& \\
-\frac{1}{x_{1}} & -\frac{y_{2}}{x_{1}x_{2}}%
\end{array}
\right)  ,\text{ \ }\tilde{U}_{2}(\lambda)=\left(
\begin{array}
[c]{cc}%
\frac{y_{2}}{x_{2}}\lambda-\left(  \frac{x_{1}y_{2}}{x_{2}}-\frac{1}{2}%
y_{1}+\frac{1}{4}\frac{x_{2}y_{2}}{x_{1}}\right)  & \lambda^{2}-x_{1}\lambda\\
& \\
-\lambda-\left(  \frac{2y_{2}^{2}}{x_{2}^{2}}+x_{1}-\frac{1}{4}\frac{x_{2}%
^{2}}{x_{1}}\right)  & -\frac{y_{2}}{x_{2}}\lambda+\left(  \frac{x_{1}y_{2}%
}{x_{2}}-\frac{1}{2}y_{1}+\frac{1}{4}\frac{x_{2}y_{2}}{x_{1}}\right)
\end{array}
\right)  \text{,}%
\]
i.e. the $g=0$ Lax representation for the system (\ref{shh}) (or
(\ref{shh1})). In a similar way, the $g(\lambda)=\lambda$ Lax representation
for the H\'{e}non-Heiles system takes the form
\begin{gather*}
L(\lambda)=%
\begin{pmatrix}
y_{1}\lambda+\frac{1}{2}x_{2}y_{2} & \lambda^{2}-x_{1}\lambda-\frac{1}{4}%
x_{2}^{2}\\
& \\
-2\lambda^{3}-2x_{1}\lambda^{2}-\left(  2x_{1}^{2}+\frac{1}{2}x_{2}%
^{2}\right)  \lambda+y_{2}^{2} & -y_{1}\lambda-\frac{1}{2}x_{2}y_{2}%
\end{pmatrix}
,\\
\\
U_{1}(\lambda)=%
\begin{pmatrix}
0 & \frac{1}{2}\\
& \\
-\lambda-2x_{1} & 0
\end{pmatrix}
,\quad U_{2}(\lambda)=%
\begin{pmatrix}
\frac{1}{2}y_{1} & \frac{1}{2}\lambda-\frac{1}{2}x_{1}\\
& \\
-\lambda^{2}-x_{1}\lambda-x_{1}^{2}-\frac{1}{2}x_{2}^{2} & -\frac{1}{2}y_{1}%
\end{pmatrix}
,
\end{gather*}
while the St\"{a}ckel transform (\ref{LU}) gives the $g(\lambda)=\lambda$ Lax
representation
\[
\tilde{L}(\lambda)=\left(
\begin{array}
[c]{cc}%
y_{1}\lambda+\frac{1}{2}x_{2}y_{2} & \lambda^{2}-x_{1}\lambda-\frac{1}{4}%
x_{2}^{2}\\
& \\
-2\lambda^{3}-2x_{1}\lambda^{2}+\left(  \frac{1}{2}x_{2}^{2}+\frac{y_{1}^{2}%
}{x_{1}}+\frac{y_{2}^{2}}{x_{1}}\right)  \lambda+y_{2}^{2} & -y_{1}%
\lambda-\frac{1}{2}x_{2}y_{2}%
\end{array}
\right)  ,
\]%
\[
\tilde{U}_{1}(\lambda)=\left(
\begin{array}
[c]{cc}%
0 & \frac{1}{2}\frac{1}{x_{1}}\\
& \\
-\frac{1}{x_{1}}\lambda-2 & 0
\end{array}
\right)  ,\text{ \ }\tilde{U}_{2}(\lambda)=\left(
\begin{array}
[c]{cc}%
\frac{1}{2}y_{2} & \frac{1}{2}\lambda-\frac{1}{2}x_{1}-\frac{1}{8}\frac
{x_{2}^{2}}{x_{1}}\\
& \\
-\lambda^{2}-\left(  x_{1}-\frac{1}{4}\frac{x_{2}^{2}}{x_{1}}\right)
\lambda+\frac{1}{2}\left(  \frac{y_{1}^{2}}{x_{1}}+\frac{y_{2}^{2}}{x_{1}%
}+x_{2}^{2}\right)  & -\frac{1}{2}y_{2}%
\end{array}
\right)
\]
for the system (\ref{shh}) (or (\ref{shh1})).
\end{example}

\begin{example}
\label{E2}(Example \ref{E1} continued). The system (\ref{e1}) has the Lax
pairs given by (\ref{14}) and (\ref{125a}) with $f(\lambda)=1$; let us now
also choose $g(\lambda)=1$. In the first step we construct the Lax pairs
$\left(  L(\lambda,\alpha),U_{j}(\lambda,\alpha)\right)  $ for the extended
system (\ref{r1}). We obtain%
\begin{equation}
L(\lambda,\alpha)=\left(
\begin{array}
[c]{cc}%
v(\lambda) & u(\lambda)\\
& \\
w(\lambda)+\alpha\left[  \frac{2\lambda^{3}}{u(\lambda)}\right]  _{+} &
-v(\lambda)
\end{array}
\right)  \label{zosia}%
\end{equation}
while $U_{j}(\lambda,\alpha)$ are then given by (\ref{125a}) with
$L(\lambda,\alpha)$ given by (\ref{zosia}). Thus,
\[
\tilde{L}(\lambda)=\left(
\begin{array}
[c]{cc}%
v(\lambda) & u(\lambda)\\
& \\
w(\lambda)+\tilde{H}_{1}\left[  \frac{2\lambda^{3}}{u(\lambda)}\right]  _{+} &
-v(\lambda)
\end{array}
\right)
\]
and%
\[
\tilde{U}_{1}(\lambda)=-\frac{1}{V_{2}^{(3)}}U_{2}(\lambda,\alpha=\tilde
{H}_{1}),\ \ \ \ \tilde{U}_{2}(\lambda)=U_{1}(\lambda,\alpha=\tilde{H}%
_{1})-\frac{V_{1}^{(3)}}{V_{2}^{(3)}}U_{2}(\lambda,\alpha=\tilde{H}_{1}),\
\]%
\[
\tilde{U}_{3}(\lambda)=U_{3}(\lambda,\alpha=\tilde{H}_{1})-\frac{V_{3}^{(3)}%
}{V_{2}^{(3)}}U_{2}(\lambda,\alpha=\tilde{H}_{1}).
\]
As in Example \ref{E1}, we present the explicit form of these formulas in the
Viet\'{e} coordinates (\ref{Viete}). For seed system, generated by
Hamiltonians (\ref{s1}), the $g(\lambda)=1$ Lax operator $L(\lambda)$ becomes
\[
L(\lambda)=\left(
\begin{array}
[c]{cc}%
-p_{{3}}{\lambda}^{2}-\left(  q_{{1}}p_{{3}}+p_{{2}}\right)  \lambda-q_{{1}%
}p_{{2}}-q_{{2}}p_{{3}}-p_{{1}} & {\lambda}^{3}+{\lambda}^{2}q_{{1}}%
+\lambda\,q_{{2}}+q_{{3}}\\
& \\
2{\lambda}^{2}-\left(  {p_{{3}}^{2}}+2q_{{1}}\right)  \lambda-q_{{1}}{p_{{3}%
}^{2}}-2\,p_{{2}}p_{{3}}+2{q_{{1}}^{2}}-2q_{{2}} & p_{{3}}{\lambda}%
^{2}+\left(  q_{{1}}p_{{3}}+p_{{2}}\right)  \lambda+q_{{1}}p_{{2}}+q_{{2}%
}p_{{3}}+p_{{1}}%
\end{array}
\right)
\]
while%
\[
U_{1}(\lambda)=\left(
\begin{array}
[c]{cc}%
0 & \frac{1}{2}\\
& \\
0 & 0
\end{array}
\right)  ,\text{ \ }U_{2}(\lambda)=\left(
\begin{array}
[c]{cc}%
-\frac{1}{2}p_{3} & \lambda+q_{1}\\
& \\
1 & \frac{1}{2}p_{3}%
\end{array}
\right)  \text{, }%
\]%
\[
U_{3}(\lambda)=\left(
\begin{array}
[c]{cc}%
-\frac{1}{2}p_{3}\lambda-\frac{1}{2}q_{1}p_{3}-\frac{1}{2}q_{2} & \frac{1}%
{2}\lambda^{2}+\frac{1}{2}q_{1}\lambda+\frac{1}{2}q_{2}\\
& \\
\lambda-\frac{1}{2}p_{3}^{2}-q_{1} & \frac{1}{2}p_{3}\lambda+\frac{1}{2}%
q_{1}p_{3}+\frac{1}{2}q_{2}%
\end{array}
\right)  .
\]
The St\"{a}ckel transform (\ref{LU})\ of the above Lax pairs yields
\[
\tilde{L}(\lambda)=\left(
\begin{array}
[c]{cc}%
-p_{{3}}{\lambda}^{2}-\left(  q_{{1}}p_{{3}}+p_{{2}}\right)  \lambda-q_{{1}%
}p_{{2}}-q_{{2}}p_{{3}}-p_{{1}} & {\lambda}^{3}+{\lambda}^{2}q_{{1}}%
+\lambda\,q_{{2}}+q_{{3}}\\
& \\%
\begin{array}
[c]{c}%
2{\lambda}^{2}-\left(  {p_{{3}}^{2}}+2q_{{1}}\right)  \lambda-\frac{2}{q_{2}%
}(\,{q_{{1}}^{2}}p_{{2}}p_{{3}}+q_{{1}}q_{2}{p_{{3}}^{2}}+q_{{1}}p_{{1}}%
p_{{3}}\\
+q_{{1}}{p_{{2}}^{2}}+q_{2}\,p_{{2}}p_{{3}}-\frac{1}{2}q_{2}q_{{3}}{p_{{3}%
}^{2}}+p_{{1}}p_{{2}}-q_{{1}}q_{{3}})
\end{array}
& p_{{3}}{\lambda}^{2}+\left(  q_{{1}}p_{{3}}+p_{{2}}\right)  \lambda+q_{{1}%
}p_{{2}}+q_{{2}}p_{{3}}+p_{{1}}%
\end{array}
\right)
\]
and%
\begin{align*}
\tilde{U}_{1}(\lambda)  &  =\left(
\begin{array}
[c]{cc}%
\frac{1}{2}\frac{p_{3}}{q_{2}} & -\frac{1}{2}\frac{1}{q_{2}}\lambda-\frac
{1}{2}\frac{q_{1}}{q_{2}}\\
& \\
-\frac{1}{q_{2}} & -\frac{1}{2}\frac{p_{3}}{q_{2}}%
\end{array}
\right)  \text{, \ \ }\tilde{U}_{2}(\lambda)=\left(
\begin{array}
[c]{cc}%
\frac{1}{2}\frac{q_{1}p_{3}}{q_{2}} & -\frac{1}{2}\frac{q_{1}}{q_{2}}%
\lambda+\frac{1}{2}-\frac{1}{2}\frac{q_{1}^{2}}{q_{2}}\\
& \\
-\frac{q_{1}}{q_{2}} & -\frac{1}{2}\frac{q_{1}p_{3}}{q_{2}}%
\end{array}
\right)  ,\\
& \\
\tilde{U}_{3}(\lambda)  &  =\left(
\begin{array}
[c]{cc}%
-\frac{1}{2}p_{3}\lambda-\frac{1}{2}q_{1}p_{3}-\frac{1}{2}p_{2}+\frac{1}%
{2}\frac{q_{3}p_{3}}{q_{2}} & \frac{1}{2}\lambda^{2}+\frac{1}{2}(q_{1}%
-\frac{q_{3}}{q_{2}})\lambda+\frac{1}{2}q_{2}-\frac{1}{2}\frac{q_{1}q_{3}%
}{q_{2}}\\
& \\
\lambda-\frac{1}{2}p_{3}^{2}-q_{1}-\frac{q_{3}}{q_{2}} & \frac{1}{2}%
p_{3}\lambda+\frac{1}{2}q_{1}p_{3}+\frac{1}{2}p_{2}-\frac{1}{2}\frac
{q_{3}p_{3}}{q_{2}}%
\end{array}
\right)  ,
\end{align*}
i.e. the respective Lax pairs for transformed system generated by Hamiltonians
(\ref{s2}).
\end{example}

\section{Acknowledgments}

MB wishes to express his gratitude for Department of Science, Link\"{o}ping,
University, Sweden, for their kind hospitality.

\end{document}